\documentclass[aps,twocolumn,groupedaddress,amsmath,amssymb]{revtex4}
\usepackage{amsmath, dsfont}
\usepackage{amssymb}
\usepackage[dvips]{graphics}
\usepackage{epsfig}
\usepackage{calc}
\usepackage{amsfonts}
\usepackage{graphicx}
\usepackage[ansinew]{inputenc}

\begin{document}
\setcounter{page}{1}
\title[]{Lifshitz black holes  with a time-dependent scalar field in Horndeski theory}

\author{Mois\'es Bravo Gaete}\email{mbravog-at-inst-mat.utalca.cl}
\affiliation{Instituto de Matem\'atica y F\'{\i}sica, Universidad de
Talca, Casilla 747, Talca, Chile.}

\author{Mokhtar Hassaine}\email{hassaine-at-inst-mat.utalca.cl}
\affiliation{Instituto de Matem\'atica y F\'{\i}sica, Universidad de
Talca, Casilla 747, Talca, Chile.}

\begin{abstract}
In arbitrary dimensions, we consider a particular  Horndeski action
given by the Einstein-Hilbert Lagrangian with a cosmological
constant term, while the source part is described by a real scalar
field with its usual kinetic term together with a nonminimal kinetic
coupling. In order to evade the no-hair theorem, we look for solutions where the radial 
component of the conserved current vanishes identically. Under this hypothesis, we prove that this model can not 
accommodate Lifshitz solutions with a radial scalar field. This problem is finally circumvented by turning on the  time 
dependence of the scalar field, and we obtain a Lifshitz black hole solution with a fixed value of the dynamical exponent $z=\frac{1}{3}$. The same metric is also shown 
to satisfy the field equations arising only from the variation of the matter source.
\end{abstract}

\maketitle

%%%%%%%%%%%%%%%%%%%%%%%
\section{Introduction}
%%%%%%%%%%%%%%%%%%%%%%%
During the last decade, there has been an intense activity to
promote the ideas underlying the gauge-gravity duality to
non-relativistic physics. The hope is to gain a better understanding
of some strongly coupled condensed matter physics phenomena observed
in laboratories, for a review see e. g. \cite{Hartnoll:2009sz}.  In
this context, the so-called Schr\"odinger or Lifshitz spacetimes are
the natural candidates to the be the gravity duals for
non-relativistic scale invariant theories,
\cite{Son:2008ye,Balasubramanian:2008dm, Kachru:2008yh}. In the
present work, we are concerned with the Lifshitz spacetimes given by
\begin{equation}
ds_{{\cal L}}^{2}=-{r^{2z}}dt^{2}+\frac{dr^2}{r^{2}}
+{r^{2}}d\vec{x}_{D-2}^2 ,\label{Lifshitz}
\end{equation}
where the dynamical exponent $z$ reflects the anisotropy of the
scaling symmetry
$$
t\to\lambda^z t,\qquad r\to\frac{r}{\lambda},\qquad
\vec{x}\to\lambda \vec{x}.
$$
Here, $\vec{x}$ denotes a $(D-2)-$dimensional vector. As it is now
well-known, for $z\not=1$, in order for the Einstein gravity to
accommodate the Lifshitz spacetime, some extra matter source is
required, like $p-$form gauge fields or some Proca model, see e. g.
\cite{Pang:2009pd}. There also exists the option of considering
higher-order gravity theories for which there exist examples of
Lifshitz black holes without source, see e. g.
\cite{AyonBeato:2009nh,AyonBeato:2010tm}. By Lifshitz black holes,
we mean a black hole geometry whose asymptotic behavior reproduces
the Lifshitz spacetime (\ref{Lifshitz}). In this work, we deal with
a source of the Einstein equations given by a real scalar field with
its usual kinetic term together with a nonminimal kinetic coupling.
More precisely, we consider the following $D-$dimensional action
{\small\begin{eqnarray}
 S=\int\sqrt{-g}d^{D}x\left(R-2\Lambda
  -\frac{1}{2}\left(  \alpha g_{\mu\nu}-\eta
G_{\mu\nu}\right)
\nabla^{\mu}\phi\nabla^{\nu}\phi\right),\label{action1}
\end{eqnarray}}
where $R$ and $G_{\mu\nu}$ stand respectively for the Ricci scalar
and the Einstein tensor. This model is part of the so-called
Horndeski action which is the most general tensor-scalar action
yielding at most to second-order field  equations in four dimensions
\cite{Horndeski:1974wa}. The action also enjoys the shifting
symmetry $\phi\to\phi+\mbox{const.}$. The field equations obtained
by varying the action with respect to the two dynamical fields
$g_{\mu\nu}$ and $\phi$ read
\begin{subequations}
\label{eqs}
\begin{eqnarray}
\label{eqmetric}
&&G_{\mu\nu}+\Lambda g_{\mu\nu}=\frac{1}2\left[{\alpha} {T}_{\mu \nu}^{(1)}+{\eta} {T}_{\mu \nu}^{(2)}\right],\\
\nonumber\\
\label{eqphi} &&\nabla_{\mu}\left[  \left(  \alpha g^{\mu\nu}-\eta
G^{\mu\nu}\right) \nabla_{\nu}\phi\right]  =0,
\end{eqnarray}
\end{subequations}
where the stress tensors $T_{\mu\nu}^{(i)}$ are defined by {\small
\begin{eqnarray} \label{stresstensor} &&T_{\mu\nu}^{(1)}
=\Big(\nabla_{\mu}\phi\nabla_{\nu}\phi-\frac
{1}{2}g_{\mu\nu}\nabla_{\lambda}\phi\nabla^{\lambda}\phi \Big).\\
&&T_{\mu\nu}^{(2)}= \frac{1}{2}\nabla_{\mu}\phi\nabla_{\nu }\phi
R-2\nabla_{\lambda}\phi\nabla_{(\mu}\phi R_{\nu)}^{\lambda}
-\nabla^{\lambda}\phi\nabla^{\rho}\phi R_{\mu\lambda\nu\rho}\nonumber\\
&&-(\nabla_{\mu}\nabla^{\lambda}\phi)(\nabla_{\nu}\nabla_{\lambda}%
\phi)+(\nabla_{\mu}\nabla_{\nu}\phi)\square\phi+\frac{1}{2}G_{\mu\nu}%
(\nabla\phi)^{2}\nonumber\\
&&-g_{\mu\nu}\left[ -\frac{1}{2}(\nabla^{\lambda}\nabla^{\rho}\phi
)(\nabla_{\lambda}\nabla_{\rho}\phi)+\frac{1}{2}(\square\phi)^{2}%
-\nabla_{\lambda}\phi\nabla_{\rho}\phi
R^{\lambda\rho}\right].\nonumber
\end{eqnarray}}
The first exact black hole solution of these equations without
cosmological constant was found in \cite{Rinaldi:2012vy}. However,
in this case, the scalar field becomes imaginary outside the
horizon. Recently, this problem has been circumvented by adding a
cosmological constant term yielding to asymptotically locally (A)dS
(and even flat for $\alpha=0$)  black hole solutions with a real
scalar field outside the horizon \cite{Anabalon:2013oea}. The field
equations (\ref{eqs}) admit other interesting solutions with a
nontrivial and regular time-dependent scalar field on a static and
spherically symmetric spacetime \cite{Babichev:2013cya}.
Interestingly enough, this solution in the particular case of
$\Lambda=\eta=0$ reduces to an unexpected stealth configuration on
the Schwarzschild metric \cite{Babichev:2013cya}.

Because of the anisotropy symmetry, the $(D-2)-$ dimensional base manifold of Lifshitz spacetimes (\ref{Lifshitz}) is planar. For this reason, we will 
restrict ourselves in looking for black hole solutions whose horizon topology is  planar. Moreover, in order to escape from the no-hair theorem established in 
\cite{Hui:2012qt}, we will also impose by hand that the radial component of the conserved current vanishes identically without restricting the radial dependence of 
the scalar field, that is
\begin{eqnarray}
\alpha g^{rr}-\eta G^{rr}=0.
\label{cond}
\end{eqnarray}
Note that in all the references previously cited \cite{Rinaldi:2012vy,Babichev:2013cya}, the different authors also consider this restriction that  simplifies 
the field equations. Under these two hypothesis (planar base manifold and the condition (\ref{cond})), we will see that the only solution with a static radial scalar field 
is a planar AdS black hole already reported in Ref. \cite{Anabalon:2013oea}. Nevertheless, in order to extend the space of admissible planar black hole solutions, we turn on the time 
dependence of the scalar field. In this case, the condition  (\ref{cond}) will impose that the time dependence of the scalar field is linear. In doing so, we will effectively obtain a Lifshitz black hole solution with a linear time-dependent scalar field for a specific value of the dynamical exponent $z=\frac{1}{3}$. The plan of the paper is organized as follows. In the next section, we provide the general analysis for a static and radial scalar field, and see that the only solution with a planar base manifold is the AdS black hole solution obtained in \cite{Anabalon:2013oea}. In Sec. III, we turn on the time dependence of the scalar field and construct a Lifshitz black hole solution with a time-dependent scalar field characterized by a dynamical exponent $z=\frac{1}{3}$. The last section is devoted to our conclusions.

%%%%%%%%%%%%%%%%%%%%%%%%%%%%%%%%%%%%%%%%%%%%%%%%%%%%%%%
\section{General analysis with a static scalar field}
%%%%%%%%%%%%%%%%%%%%%%%%%%%%%%%%%%%%%%%%%%%%%%%%%%%%%%%

Let us consider the following Ansatz
\begin{equation}\label{metric}
ds_{D,\gamma}^{2}=-h(r) dt^{2}+\frac{dt^{2}}{f(r)}+r^{2} d
\Omega^{2}_{D-2,\gamma},\quad \phi=\phi(r)
\end{equation}
where $ d \Omega^{2}_{D-2,\gamma}$ represents the line element of a
$(D-2)-$dimensional sphere, plane or hyperboloid which corresponds respectively to $\gamma=1, 0$ or $\gamma=-1$. Here, we are mainly interested on the planar 
case $\gamma=0$ but we prefer to keep this general form in order to stress the particularity of considering the planar case $\gamma=0$. In this case, the condition on the radial component of the current conservation (\ref{cond}) permits to relate the two metric functions as 
\begin{equation}\label{relfh}
f =\frac { h \left[2\,\alpha\,r^{2}+(D-3)(D-2)\gamma\,\eta\right] }{
\left( D-2 \right) \eta\, \left[ h'r+ \left( D-3 \right) h  \right]
}.
\end{equation}
We are now in position to solve the Einstein equations (\ref{eqmetric}); their $(r,r)$ component 
allows to express $\psi_{\tiny{\mbox{static}}}:=\phi'$ as
\begin{eqnarray*}\label{phistatic}
\psi(r)^{2}_{\tiny{\mbox{static}}}=-{\frac {4\,{r}^{2} \left(
\Lambda\,\eta+\alpha \right) \left( D-2 \right)  \left[ r h' +h
\left( D-3 \right) \right] }{ \left[ 2\,\alpha\,{r}^{2}+ \left( D-2
\right)  \left( D-3 \right) \eta\,\gamma \right] ^{2}h }}.
\end{eqnarray*}
Substituting these two expressions in the remaining independent
Einstein equations, that is the $(t,t)$ or $(i,i)$ component, one
obtains a second order differential equation for the metric function
$h$. Under the following substitution 
\begin{eqnarray}\label{metricfunh}
h(r)=-\frac{\mu}{r^{D-3}}+\frac{2}{r^{D-3}}\,\int
\frac{j(r)\,dr}{2\alpha r^{2}+\gamma\eta(D-2)(D-3)},
\end{eqnarray}
where $\mu$ is an integration constant, the differential equation
becomes a third-order algebraic equation for the function $j(r)$,
\begin{eqnarray}\label{mastereqstatic}
\varepsilon_{\tiny{\mbox{static}}}&:=&j \left[ {r}^{2} \left(
\Lambda\,\eta-\alpha \right) -\gamma\,\eta \left( D-2 \right) \left(
D-3 \right) \right]\nonumber\\
&+&{\frac {C_{{0}} j^{3/2}}{{ r}^{\frac{D-4}{2}}}}=0,
\end{eqnarray}
where $C_{0}$ is a second integration constant. From this last expression, it is clear that in the case $\gamma=0$, the only possibility for the function $j$, 
apart from the trivial case $j=0$ which is of little interest, is for $j\propto r^{D}$ whose full integration yields the planar AdS black hole solution reported previously in 
\cite{Anabalon:2013oea}. It is also interesting to stress from the expression (\ref{mastereqstatic}) that in the planar case $\gamma=0$, the point $\alpha=\eta\Lambda$ is degenerate 
in the sense that it will impose $C_0=0$, and in turn, the function $j$ will be undetermined. Hence, for $\gamma=0$ and for $\alpha=\eta\Lambda$, any metric functions $f$ and $h$ satisfying the constraint (\ref{relfh}) will be solution of the field equations. 

In what follows, we will see that a way in order to extend the space of admissible planar black hole solutions is to turn on the time dependence of the scalar field.

%%%%%%%%%%%%%%%%%%%%%%%%%%%%%%%%%%%%%%%%%%%%%%%%%%%%%%%%%%%%%%%%
\section{Lifshitz black hole with a time-dependent scalar field}
%%%%%%%%%%%%%%%%%%%%%%%%%%%%%%%%%%%%%%%%%%%%%%%%%%%%%%%%%%%%%%%%
As seen previously, Lifshitz spacetimes (\ref{Lifshitz}) with a static radial scalar field can not source the particular Horndeski considered here (\ref{action1}).  Inspired by the work done in \cite{Babichev:2013cya}, we wonder
wether there exist Lifshitz static black hole solutions with a
nontrivial time-dependent scalar field $\phi=\phi(t,r)$ satisfying
the field equations (\ref{eqs}). In doing so, we consider again the same Ansatz for the metric (\ref{metric}). We firstly note that the $(t,r)$ component of the
Einstein equations gives
\begin{eqnarray}\label{trcomponent}
&&\Big\{2\, \left( D-2 \right) \eta\,\dot{\phi}'\,f\,h \, r
-\dot{\phi} \Big[\Big( \eta\, ( D-2 ) \left( D-3 \right) ( f -\gamma
)\nonumber\\
&&-2\,\alpha\,{r}^{2} \Big)\,h + ( D-2 )\eta\,r\,h'\,f
\Big]\Big\}\,\phi'=0,
\end{eqnarray}
where $( \dot{} )$ denotes the derivative with respect to the time
$t$ and $( ' )$ the derivative with respect to the radial
coordinate. Apart from the trivial option $\phi^{\prime}=0$ that
does not yield interesting result, this equation is easily
integrated as
\begin{eqnarray}\label{genphit}
\phi(t,r)=\zeta(r)+q(t)\,e^{\chi(r)},
\end{eqnarray}
where $\zeta$ (resp. $q$) is a function of the radial coordinate
(resp. of the time), and where we have defined
\begin{eqnarray}\label{eqchi}
\chi(r)&=&\frac{1}{2}\,\int \Big[{\frac {\eta\, ( D-2 )  ( D-3 )
\left(f-\gamma \right) -2\,\alpha\,{r}^{2}}{\left( D-2
\right) \eta\,f\,r}}\nonumber\\
&+&\frac{h'}{h}\Big]\,dr.
\end{eqnarray}
One can see that that under our hypothesis  (\ref{relfh}), the  expression between
the brackets (\ref{eqchi}) vanishes, yielding a scalar field to be given by
\begin{eqnarray*}
\phi(t,r)=\zeta(r)+q(t).
\end{eqnarray*}
Injecting this expression into the conservation equation (\ref{eqphi}), this  implies that the scalar field must be linear in time 
\begin{eqnarray}\label{scalarfieldt}
\phi(t,r)=\zeta(r)+\phi_{1}\,t,
\end{eqnarray}
where $\phi_1$ is an integration constant. The $(r,r)$ component of
the Einstein equations (\ref{eqmetric}) allows to express
$\varphi:=\zeta'$ as
\begin{eqnarray}
\varphi(r)^{2}=\frac { \left( D-2 \right) \eta\, {\phi_1}^{2}\, h'
\,r }{ \left[ 2\,\alpha\,{r}^{2}+ \left( D-2 \right)  \left( D-3
\right) \gamma \,\eta\right]
h^{2}}+\psi(r)^{2}_{\tiny{\mbox{static}}},
\end{eqnarray}
and the remaining independent Einstein equation, given by the
$(t,t)$ or $(i,i)$ component, yields a second order differential
equation for the metric function $h$. As before, through the substitution
(\ref{metricfunh}), the metric function $h$ will be given by (\ref{metricfunh}) where now $j$ is a solution of the following third-order algebraic equation
\begin{eqnarray}\label{mastereqtdependent}
\frac{1}{8}\,\left( D-3 \right)\,\eta\,{\phi_{1}}^{2} \,{\left[
\left( D-2 \right) \left( D-3 \right) \gamma\,\eta+2\,
\alpha\,{r}^{2} \right] ^{2}{r}^{D-4}}&&
\nonumber\\
+\varepsilon_{\tiny{\mbox{static}}}=0,&&
\end{eqnarray}
where $\varepsilon_{\tiny{\mbox{static}}}$ is defined by (\ref{mastereqstatic}).

Here, it is interesting to note that, in contrast with the purely static case, the point
defined by $\alpha=\eta\,\Lambda$ and $\gamma=0$ is not degenerate if one considers time-dependent scalar field, $\phi_1\not=0$. Indeed, in this case, we obtain a Lifshitz black hole solution 
with a dynamical exponent $z=\frac{1}{3}$ given by
\begin{subequations}
\begin{eqnarray}
\label{m1}
&&ds^2=-r^{\frac{2}{3}}g(r)dt^2+\frac{dr^2}{r^2\, g(r)}+r^2d\vec{x}_{D-2}^2,\\
\label{m2}
&&g(r)=1-\frac{M}{r^{\frac{3\,D-7}{3}}},\\
&&\phi(t,r)=\int \varphi(r) \,dr+\phi_{1}\,t, 
\label{s1}
\end{eqnarray}
\end{subequations}
where
\begin{eqnarray}
\varphi(r)^2= \frac{1}{r^{\frac{8}{3}}f(r)}\,\left[{\frac
{{\phi_{{1}}}^{2}}{ f \left( r \right)}}-{\frac {3\,{\phi_{{1}}}^{
2} \left( D-3 \right) }{ \left( 3\,D-7 \right)
}}-\frac{4\,r^{\frac{2}{3}}}{\eta}\right],
\end{eqnarray}
and, where the coupling constants are tied as
\begin{eqnarray}\label{alphat}
\alpha=\frac{1}{6}\, \left( D-2 \right) \eta\, \left( 3\,D-7
\right),\qquad \Lambda=\frac{\alpha}{\eta}.
\end{eqnarray}
Two remarks can be made concerning this solution. Firstly, the limiting case $M=0$ is well-defined and hence the
field equations (\ref{eqs}) may accommodate pure Lifshitz spacetimes
(\ref{Lifshitz}) only with a time-dependent scalar field. It is also interesting to note that the same metric (\ref{m1}-\ref{m2}) is also a particular solution of the Horndeski equations 
(\ref{{eqs}}) without the Einstein-Hilbert-$\Lambda$ pieces, that is it satisfies the equations
\begin{eqnarray}
{\alpha}{T}_{\mu \nu}^{(1)}+{\eta}{T}_{\mu \nu}^{(2)}=0,
\label{stealth}
\end{eqnarray}
provided that the scalar field is given by
\begin{eqnarray}
&&\phi(t,r)=\phi_{1}\left(\int \varphi(r)dr+t\right)\\
&&\varphi(r)=\pm\,\frac{1}{f \left( r \right)\,
{r}^{\frac{4}{3}}}\,\sqrt {{\frac {\big[\left(3\,D-7\right)- 3\left(
D-3 \right) f \left( r \right) \big]}{\big(3\,D-7\big)}}},\nonumber
\end{eqnarray}
and for the coupling constants given by (\ref{alphat}).

%%%%%%%%%%%%%%%%%%%%%%
\section{Conclusions}
%%%%%%%%%%%%%%%%%%%%%%
Here, we have considered a particular case of the Horndeski theory
whose gravity theory is given by the Einstein piece and, whose
matter source is described by a scalar field with its usual kinetic
term as well as an additional nonminimal kinetic coupling. For this
model and for a static scalar field, we have shown that besides a planar AdS black hole, this system can 
not accommodate other solutions with a planar base manifold. In order to circumvent this problem, we have seen that turning on the time dependence
 of the scalar field is primordial  to obtain Lifshitz solutions. We have effectively derived a Lifshitz black hole solution for a particular 
 value of the dynamical exponent $z=\frac{1}{3}$. There is a priori no physical reasons to explain the occurrence of this particular value of the dynamical exponent. It 
 will be interesting  to see wether others sectors of the Horndeski action may accommodate more general Lifshitz solutions with others values of the dynamical exponents.

We also  believe that, because of structure of the solutions obtained in
this paper as well as those derived in
\cite{Rinaldi:2012vy,Anabalon:2013oea,Babichev:2013cya}, the model
considered here or maybe more general sectors of the Horndeski
theory can be a good laboratory in order to gain more insight
concerning the black hole solutions with scalar field. The task of
finding interesting solutions can also be explored for the recent
theory  formulated in \cite{Padilla:2013jza} and involving more than
one scalar field but still yielding to second-order field equations.

%%%%%%%%%%%%%%%%%%%%%%%%%%%%%%%%%%%%%%%%%%%%%%%%%%%%%%%%%%%%%%%%%%%%%%%%%%%%%%%%%%%%%%%%%%%%%%%%
\begin{acknowledgments}
We thank Julio Oliva for useful discussions. MB is supported by BECA
DOCTORAL CONICYT 21120271. MH was partially supported by grant
1130423 from FONDECYT, by grant ACT 56 from CONICYT and from
CONICYT, Departamento de Relaciones Internacionales ``Programa
Regional MATHAMSUD 13 MATH-05''.
\end{acknowledgments}
%%%%%%%%%%%%%%%%%%%%%%%%%%%%%%%%%%%%%%%%%%%%%%%%%%%%%%%%%%%%%%%%%%%%%%%%%%%%%%%%%%%%%%%%%%%%%%%%%%

%%%%%%%%%%%%%%%%%%%%%%%%%%%

\end{document}